\documentclass{collective-intelligence}

\usepackage{graphicx}
\usepackage{colortbl}              

\begin{document}
\bibliographystyle{agsm}

\title{COLLECTIVE COGNITIVE AUTHORITY: \\
EXPERTISE LOCATION VIA SOCIAL LABELING}
%
%
%
%
%

\numberofauthors{1} 
%
\author{
%
%
\alignauthor
Terrell G. Russell\\
       \affaddr{University of North Carolina at Chapel Hill}\\
       \affaddr{100 Manning Hall}\\
       \affaddr{Chapel Hill, NC}\\
       \email{unc@terrellrussell.com}
}

\maketitle
\begin{abstract}
The problem of knowing who knows what is multi-faceted.  Knowledge and expertise lie on a spectrum and one's expertise in one topic area may have little bearing on one's knowledge in a disparate topic area.  In addition, we continue to learn new things over time.  Each of us see but a sliver of our acquaintances' and co-workers' areas of expertise.  By making explicit and visible many individual perceptions of cognitive authority, this work shows that a group can know what its members know about in a relatively efficient and inexpensive manner.
\end{abstract}

\section{Introduction}
Cognitive authority is the foil to administrative authority \cite{Wilson1983Second-hand-kno}.  Administrative authority is that which one has through rank or position.  Cognitive authority is that which is granted to you by others because of what they think you know about.

Cognitive authority is a subjective measurement and should be respected as such.  There are no right answers to questions of cognitive authority, although, when taken collectively, an assessment of it can be seen as a barometer of one's standing among peers.

Making a collective assessment visible, bringing the tacit individual knowledge into the realm of the explicit, and performing a sanity check on that assessment is the thrust of this paper.  This work shows that a group's evaluations of an individual's areas of expertise can be gathered and potentially serve as useful loose credentials; loose credentials that may be useful when more expensive or heavyweight reputation cues may not be viable.

\section{Background}

We satisfice; we satisfy with what is sufficient \cite{Simon1957Models-of-Man:-Socia}. We use what information we have to make decisions that we deem to be good enough at the time. We often seek out more information before making a decision but we have, what Simon called, ``bounded rationality.'' We have imperfect information, limited attention and money, limited processing power and limited time, but we still need to make decisions.

Choo's Decision Behavior Model shows us that contextualized decision making happens within organizations based on cognitive limits, information quality and availability, and the values of the organization \cite{Choo1996The-knowing-organiza}.  These inputs are handled with bounded rationality and within the confines of performance concerns, and whether the decision is good enough, among other simplifications.  This decision making behavior is both rationally expected and observed.

Even knowing we will never have perfect information when working in these limited environments, we can arguably make better decisions if we can improve or increase the amount of information on hand when making decisions.  Having more good information reduces uncertainty about the environment surrounding a decision, but it does not necessarily reduce equivocality.  To reduce equivocality, or ambiguity, of the information we have on hand, we need sensemaking and a perspective that comes from ``retrospective interpretations'' of earlier data and decisions \cite{Choo1996The-knowing-organiza}.  We need to have \emph{seen this before} and \emph{know what it means}.  What we need to make good decisions, in addition to good information, is called expertise.

There is a vast amount of latent, untapped information in the environment around us. Some of it is in the built world, some of it is in the natural world (too big, too small, hidden in non-visible wavelengths, etc.), and some of it is in the heads of those around us. Cross  noted that 85\% of managers immediately mentioned specific people when asked ``to describe sources of information important to successful completion of their project'' \cite{Cross2004More-Than-an-Answer:}.  They went on to write:

\begin{quote}
As one manager said, ``I mean the whole game is just being the person that can get the client what they need with [the Firm's] resources behind you. This almost always seems to mean knowing who knows what and figuring out a way to bring them to your client's issue" (R6). Very few of the named people were simply organizationally designated ``experts"; most were described as partners in information relationships.
\end{quote}

If we are informed by the right people before making decisions, and they help us decide what we are looking for \cite{Belkin1982ASK-for-Information-}, then we may improve our knowledge and understanding of a situation or problem at the time when we need to decide. \textbf{Knowing from whom we should get our information, when we are not sure of what we need, is a hard problem.}

Expertise location, for this reason, has been a focus of the knowledge management field for many years. Knowledge management has also focused on the process of organizational learning and dissemination of that learning within the organization. In many cases, this has been done through the tracking of created documents and other knowledge artifacts \cite{Martin2008Knowledge-Management}.

An additional approach should consist of uncovering that which has not \emph{yet} been recorded -- that information which is in the heads of a group's membership. We should be equipped to hold up a mirror to help reflect an organization's insights and expertise back on itself. We need to help uncover the dark corners where we are not sure about the expertise in the room. With a regimen of self-reflection, iterated over time, I hope this problem can be made less hard. I think we can discover \emph{whom to ask} for the relatively low cost of a little sustained individual effort and some focused record-keeping in the distributed network.

\section{Study Design}
Because cognitive authority is a subjective measurement, there is no objective way to measure its ``precision'', no yard stick by which to measure its correctness.  However, if, with repeated exposure and more familiarity, a group's assessment and the assessment of the individual they are assessing become more similar over time, then I argue that this is a signal of the collective assessment's relative validity.  This validity is what makes the group's opinion matter.  This validity is what gives a group's opinion its weight.

\begin{figure}
\centering
\includegraphics[width=3.2in]{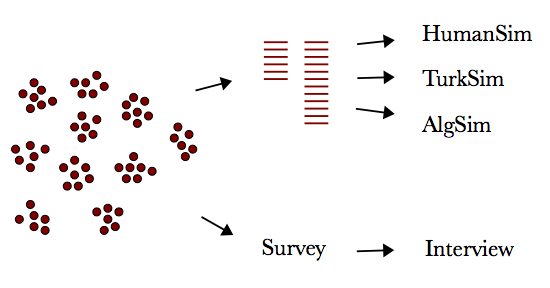}
\caption{Study design.  10 groups, their Self and Group lists about each participant, the three types of similarity ratings.  There was also a survey and a set of interviews which captured context and sentiment.}
\label{studydesign}
\end{figure}

This study had 10 different groups of individuals, mostly coworkers, use free text keywords to label each others' areas of expertise (Figure \ref{studydesign}).  The participating groups consisted of members from a family retail business, a dentist's office, two distributed software development groups, a museum education staff, a writer's network, a legal non-profit, a global engineering firm, an academic faculty group, and an academic administrative office.

Results were shared back into the group and made visible, and the process was repeated for up to five total rounds.  The resulting product was an aggregated, weighted list of words associated with each person's areas of expertise.  Inspired by the distribution of keywords affiliated with individual URLs at delicious.com, each individual's weighted list can be viewed as a specific fingerprint in the multidimensional space created by all possible keywords.

While individuals were labeling each other during each round, they were also labeling their own areas of expertise.  With each passing round, ala the Delphi model \cite{Dalkey1963An-experimental-appl,Rohrbaugh1979Improving-the-qualit,Stewart1987The-Delphi-Technique,Luo2009Delphi-Studies}, the participants were presented with what their group had said about them and had the opportunity to update (by adding, removing, or abstaining) their list of keywords about their own areas of expertise.

The original Delphi study was run in the 1950s and 1960s by the RAND corporation to help the US Government determine the nuclear capabilities of the Soviet Union \cite{Helmer1959On-the-Epistemology-,Dalkey1963An-experimental-appl}.  They were studying the unknown military futures market by asking a variety of experts to answer a battery of questions.  The answers were collated and then distributed back to the experts for additional rounds of answering the same questions - but critically, with the collective opinions of the other experts to aid their synthesis.

Rowe and Wright write that, ``in particular, the structure of the technique is intended to allow access to the positive attributes of interacting groups (knowledge from a variety of sources, creative synthesis, etc.), while pre-empting their negative aspects (attributable to social, personal and political conflicts, etc.)" \cite{Rowe1999The-Delphi-technique}.  Over the following four decades, the Delphi method has been refined and used in many other areas besides military futures, including social science predictions \cite{Linstone1975The-Delphi-Method:-T,Rowe2005Judgment-change-duri,Hsu2007The-Delphi-Technique}.

Most research has suggested that with proper preparation and consideration for expert subjects, questionnaires, and evaluation, a Delphi study can run from three to five rounds, with four being the most common number of iterations \cite{Hsu2007The-Delphi-Technique}.  Some prior Delphi studies have used post-task surveys to sample participants' reactions - from satisfaction \cite{Van-De-Van1974The-Effectiveness-of} to confidence \cite{Scheibe1975Experiments-in-Delph,Boje1982Group-Confidence-Pre} to difficulty and enjoyableness \cite{Rohrbaugh1979Improving-the-qualit}.  I employed some of the same types of questions here, especially considering the subjects were being asked to formalize their informal knowledge about one another.

A traditional Delphi study involves 1) an objective facilitator who gives ``controlled feedback'' in the aggregate, 2) a collection of independent experts in a domain (anonymous, to each other), and 3) a series of evaluations (iterations) designed to have the collective opinion of the experts predict the future in that particular domain \cite{Rowe1999The-Delphi-technique}.


In this study, the co-workers are the experts, their labels are handled by the software and not attributed to any individual labeler, and there are 5 rounds of reflection and labeling.


\begin{figure}[h]
\centering
\includegraphics[width=3.2in]{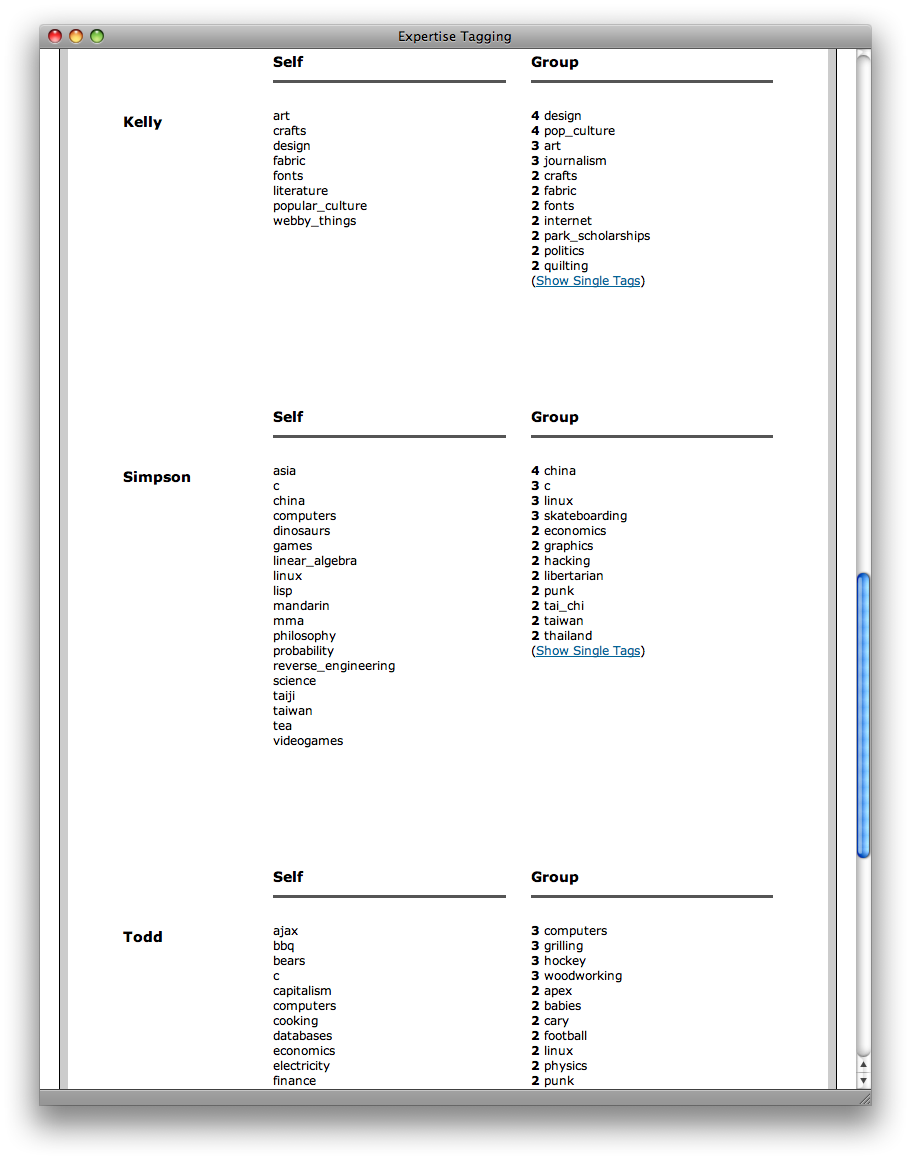}
\caption{Sample results from the labeling exercise demonstrating the Self and Group lists.}
\label{selfgroup}
\end{figure}

Data collection for this study was done primarily through a custom web interface that guided the participants through three stages per round, for up to five rounds.  The three stages included reflection on the existing labels in the system (see an example in Figure \ref{selfgroup}), a chance to add, update, or delete labels about their own areas of expertise (Self), and a chance to do the same for each of their peers (Group).

Regarding the inherent ``fuzziness'' of this type of data, Latour, Woolgar, and Nelson suggest to us that where there is a lack of contention, a social fact will be defined \cite{Latour1986Laboratory-Life:-The,Nelson1993Epistemological-Comm}.  Social tagging phenomena have demonstrated a stabilization of tagging behavior over time \cite{Russell2006Cloudalicious:-Watch,Golder2005The-Structure-of-Col}. Together, these suggest the first hypothesis:

\newtheorem{hypothesis}{Hypothesis}
\begin{hypothesis}
As the social fact of what a person knows is collectively molded by the group, a consensus will appear and converge.
\end{hypothesis}

The similarity of a group's opinion and an individual's opinion will increase over time.  That is, a similarity rating comparing two groups of words will increase from round to round of labeling.

Furthermore, the warranting principle suggests that we give more credence to information provided by others, rather than information within the control of a particular other  \cite{Walther2002Cues-Filtered-Out-Cu,Walther2009Self-Generated-Versu}.  Online or offline, information that is known to be easily manipulated is less trusted.  Additionally, Delphi-style studies increase the confidence levels of the participants \cite{Rowe2005Judgment-change-duri}. This suggests the second hypothesis: 

\begin{hypothesis}
Group members will have confidence in this system and exhibit increased trust in one another.
\end{hypothesis}

\section{Results}

Similarity ratings were generated in three distinct ways: Human coded (gold standard), Amazon's Mechanical Turk (for scale), and an Algorithmic (automated) solution based on a bag-of-words assumption.  These three methods, along with some survey results, are described here.

\subsection{Trained Human Similarity (HumanSim)}

The HumanSim ratings were designed to serve as a sanity check for this type of data collection and analysis.  Six trained raters coded a sample of the entire dataset and made 2348 comparisons of sets of words and rated their similarity on a 1-7 Likert scale with 1 meaning very dissimilar and 7 meaning nearly identical.

\begin{figure}[h]
\centering
\includegraphics[width=3.2in]{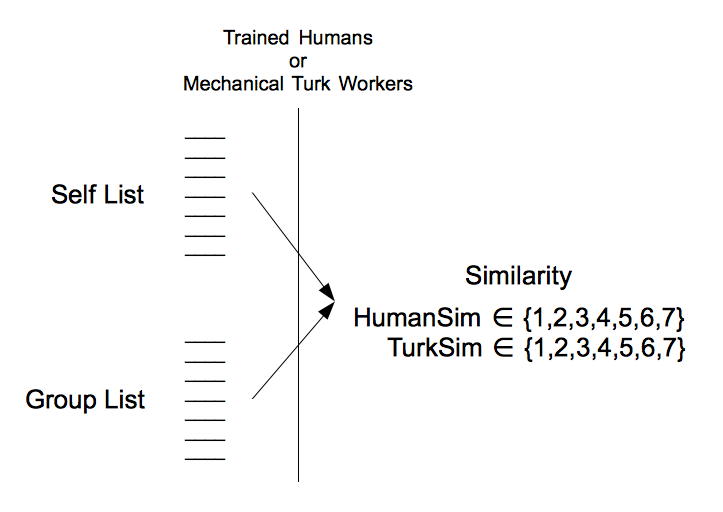}
\caption{Similarity Rating Flow: For both the HumanSim and TurkSim (next section) ratings, a Self list and a Group list were compared against one another and rated on a 1-7 scale with a high rating of 7.}
\label{humanandturkflow}
\end{figure}

They compared both weighted and unweighted lists and evaluated both random pairings (a Self list from one person with a Group evaluation of a different person) and lists ``belonging'' to a study participant (where the Self and Group lists were about that participant).  Each comparison was evaluated in an average of 15 seconds.

\begin{figure}[h]
\centering
\includegraphics[width=3.2in]{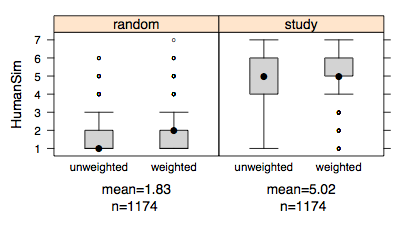}
\caption{HumanSim: Trained humans determined that random pairings of Self and Group were dissimilar and that the Self and Group lists for the same individual were similar.  When the labels were weighted, the similarity ratings were more tightly clustered at the high end of the scale.}
\label{humansim}
\end{figure}

A two-way analysis of variance (see Figure \ref{humansim}) illustrates the significance of both independent variables.  There were no significant interaction effects.  The main effect between the random design and study design was significant with a p-value of 0.000.  The main effect for weightedness was significant at the 0.01 level with a p-value of 0.0013.

Trained humans determined that labels attributed to one's areas of expertise by one's peers are similar to the labels given by someone about their own areas of expertise.  The next two methods of generating ratings are an attempt to do the same thing, but cheaper and more quickly.

\subsection{Untrained Mechanical Turk Similarity (TurkSim)}

The second of the three methods of calculating similarity of the study data was via Amazon's Mechanical Turk.  This method of analysis evaluated a series of 8773 Human Intelligence Tasks (HITs) for a total cost of \$219.33.

Similar to the HumanSim comparisons, the Turkers were asked to rate the similarity of two sets of words on a 1-7 Likert scale.  A rating of 7 meant they felt the two lists were extremely similar and conveyed the same information.  The Turkers rated data from each member of each group from each of the rounds.

\begin{figure}[h]
\centering
\includegraphics[width=3.2in]{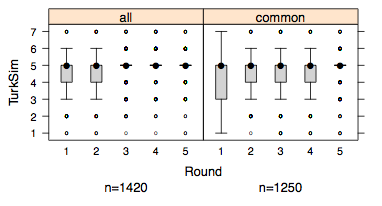}
\caption{TurkSim: Untrained Mechanical Turk workers evaluated the two sets of labels about study participants' areas of expertise to be similar (average rating of 5) with the range tightening noticeably as the rounds continued.  The ranges collapsed more quickly (Round 2 vs. Round 5) when all the words were shown, rather than only the words that appeared in the lists multiple times (``common'' words).}
\label{turksim}
\end{figure}

When looking at the similarity ratings across the entire study (see Figure \ref{turksim}), it is clear the agreement that the lists convey similar information, but there is not a dramatic rise in that agreement over time.  The range of similarity scores tightened around the mean as the rounds progressed.  This does show consensus-building, but not an increase in the raw similarity of the data being evaluated.

Untrained Mechanical Turk workers determined that labels attributed to one's areas of expertise by one's peers are similar to the labels given by someone about their own areas of expertise.  However, that similarity did not measurably increase over time.

\subsection{Algorithmic Similarity (AlgSim)}

The third method for evaluating similarity of the study data was automated and tested here to see whether it produced reproducible results, similar to that of our HumanSim gold standard and TurkSim methods.

\begin{figure}[h]
\centering
\includegraphics[width=3.2in]{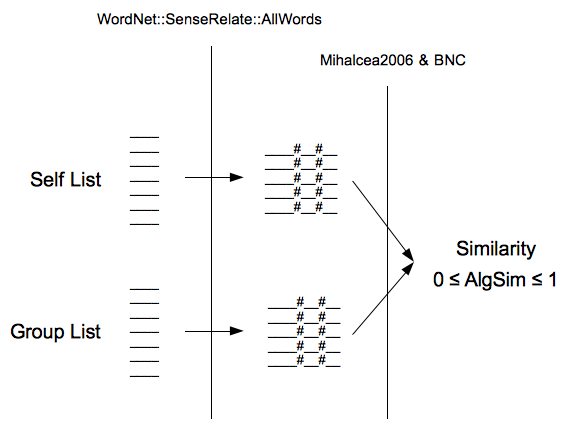}
\caption{Similarity Rating Flow: For the AlgSim ratings, both the Self list and the Group list were ``sense disambiguated'' by using a package of the WordNet project which determines which version of each word is being used (i.e. bank: money vs river).  Then each list is compared as a group against the other and rated by the algorithm in Equation \ref{eq:mihalcea_similarity}.}
\label{algsimflow}
\end{figure}

The ratings in this section were computationally generated based on an algorithm (Equation \ref{eq:mihalcea_similarity}) originally defined to calculate the similarity of sentences in the English language \cite{Mihalcea2006Corpus-based-and-kno}.  It uses a naive bag-of-words approach and ignores context such as word order.  This naivety is a positive for this analysis since there is no implicit word order in the sets of labels being evaluated.  However, the labels are related in that they are part of the same set and this algorithm allows a straightforward approach to generating a value for that relationship.  The resulting similarity scores were in the range [$0..1$] and cannot be directly compared to the human-generated Likert scale scores of 1-7 from the prior two sections.  The original lists of labels were \emph{sense disambiguated} using the WordNet database \cite{Pedersen2004WordNet::Similarity-,Pedersen2009WordNet::SenseRelate} and then compared against one another.

\begin{equation}
\resizebox{.9\hsize}{!}{$
  AlgSim(A,B) = \frac{1}{2} \left ( \frac{ \displaystyle \sum_{w \in \{A\}} (maxSim(w,B) * idf(w)) } { \displaystyle \sum_{w \in \{A\}} idf(w) } + \frac{ \displaystyle \sum_{w \in \{B\}} (maxSim(w,A) * idf(w)) } { \displaystyle \sum_{w \in \{B\}} idf(w) } \right ) 
$}
  \label{eq:mihalcea_similarity}
\end{equation}

The algorithm took each word in set $A$ and found the most similar word in set $B$ (represented by $maxSim(w,B)$) and then multiplied by the information content of that word (represented by $idf(w)$).  This summation was normalized across the information content of the entire list ($\sum_{w \in \{A\}} idf(w)$).  After each list was compared one to the other, the similarity values were averaged for the final \emph{AlgSim} value for two lists $A$ and $B$.

Using this approach, the same analysis that was performed by the Mechanical Turk workers could be duplicated much more efficiently.  The results are presented here.

\begin{figure}[h]
\centering
\includegraphics[width=3.2in]{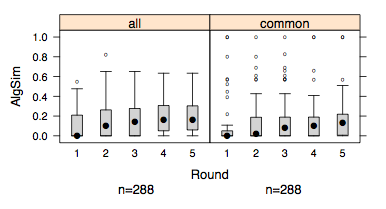}
\caption{AlgSim: The automatic algorithm detected a greater similarity in the later rounds than in the earlier rounds.  It produced a tighter range of scores when it had more information to use (when using all the words rather than just the words used more than once in each list).}
\label{algsim}
\end{figure}

Using this method the scores appear low on an absolute scale, but should be interpreted relatively.  There was a clear rise in similarity score from Round 1 to Round 5 (see Figure \ref{algsim}) for both evaluation techniques (using all the words and only the words which occurred more than once).

\begin{table}[h]
\centering
\caption{AlgSim: Repeated measures ANOVA by Round}
\begin{tabular}{lrrrrr}
  \hline
 & Df & Sum Sq & Mean Sq & F value & Pr($>$F) \\ 
  \hline
Round & 1 & 0.32 & 0.32 & 10.22 & 0.0015 \\ 
  Residuals & 286 & 8.85 & 0.03 &  &  \\ 
   \hline
\end{tabular}
\label{algsimanova-main}
\end{table}

The repeated measures analysis of variance (Table \ref{algsimanova-main}) of the AlgSim scores across time (represented by Round) for the case where all words were evaluated showed a statistically significant main effect with an alpha of 0.01 and p-value of 0.0015.  

\begin{table}[h]
\centering
\caption{AlgSim: Post-hoc ANOVA p-values by Round}
\begin{tabular}{l|cccc}
        & Round 1 & Round 2 & Round 3  & Round 4 \\
\hline
Round 2 & \textbf{0.0249*} & \cellcolor[gray]{0.8}-  & \cellcolor[gray]{0.8}-  & \cellcolor[gray]{0.8}-  \\
Round 3 &   \textbf{0.0058*}     & 0.6428  & \cellcolor[gray]{0.8}-   & \cellcolor[gray]{0.8}-    \\
Round 4 &   \textbf{0.0016**}     &   0.3906      & 0.6783   & \cellcolor[gray]{0.8}-   \\
Round 5 &   \textbf{0.0014**}    &    0.3480     &    0.6062      & 0.9072   \\
\hline
\end{tabular}
\label{algsimanova-posthoc}
\end{table}

However, looking at round by round post-hoc ANOVA analysis when all the words are used (Table \ref{algsimanova-posthoc}), the only significant differences appear to occur after the original feedback loop between Rounds 1 and 2 (p-values less than 0.05) when the participants are initially faced with their group members' feedback and labels.  No other single round or span of rounds show an effect.  This same analysis did not produce significant effects when only the words appearing more than once were used.

\subsection{Survey}

The survey accompanying this study consisted of 54 questions and was completed by 56 of the 64 participants across all 10 groups.

The survey results listed in Table \ref{surveyresults} represent the 11 original questions as well as the aggregated results from the seven included scales (see Table \ref{surveyscales}).

The scores are on a 1-7 Likert scale representing agreement: 1=Extremely Disagree, 4=Neutral, 7=Extremely Agree.

\begin{table}[h]
  \caption{Survey: Items from Selected Scales}
\resizebox{\hsize}{!}{
\centering
\begin{tabular}{|p{4.5in}|}
\hline
\textbf{Result Demonstrability} \cite{Moore1991Development-of-an-In}\\
\hline
- I would have no difficulty telling others about the results of using this system.\\
\hline
- I believe I could communicate to others the consequences of using this system.\\
\hline
- The results of using this system are apparent to me.\\
\hline
- I would have difficulty explaining why using this system may or may not be beneficial. (reverse coded)\\
\hline
\textbf{Relative Advantage} \cite{Moore1991Development-of-an-In} \\
\hline
- Using this system would enable me to accomplish tasks more quickly.\\
\hline
- Using this system would improve the quality of work I do.\\
\hline
- Using this system would make it easier to do my job.\\
\hline
- Using this system would enhance my effectiveness on the job.\\
\hline
- Using this system would give me greater control over my work.\\
\hline
\textbf{Performance Expectancy} \cite{Venkatesh2003User-Acceptance-of-I}\\
\hline
- I would find this system useful in my job.\\
\hline
- Using this system enables me to accomplish tasks more quickly.\\
\hline
- Using this system increases my productivity.\\
\hline
\textbf{Effort Expectancy} \cite{Venkatesh2003User-Acceptance-of-I} \\
\hline
- My interaction with this system would be clear and understandable.\\
\hline
- It would be easy for me to become skillful at using this system.\\
\hline
- I would find this system easy to use.\\
\hline
- Learning to operate this system would be easy for me.\\
\hline
\textbf{Facilitating Conditions} \cite{Venkatesh2003User-Acceptance-of-I}\\
\hline
- I have the resources necessary to use this system.\\
\hline
- I have the knowledge necessary to use this system.\\
\hline
- This system is not compatible with other systems I use. (reverse coded)\\
\hline
\textbf{Anxiety} \cite{Venkatesh2003User-Acceptance-of-I} \\
\hline
- I feel apprehensive about using this system.\\
\hline
- It scares me to think that I could lose a lot of information using this system by hitting the wrong key.\\
\hline
- I hesitate to use this system for fear of making mistakes I cannot correct.\\
\hline
- This system is somewhat intimidating to me.\\
\hline
\textbf{Data Quality} \cite{Wang1996Beyond-Accuracy:-Wha} \\
\hline
- This system produced data in conformance with the actual or true values.\\
\hline
- This system produced data that is applicable and relevant to my job.\\
\hline
- This system produced data that is intelligible and clear.\\
\hline
- This system produced data that is easily accessible.\\
\hline
\end{tabular}
}
\label{surveyscales}
\end{table}

\begin{table}[h]
  \caption{Survey Scales and Ratings}
  \resizebox{\hsize}{!}{
\centering
\begin{tabular}{p{2.5in} m{.5in}}
\hline
Original Items & Average Rating \\
\hline
I am comfortable with my group's tags about my areas of expertise. & 5.439 \\
I am happy with my group's tags about my areas of expertise. & 5.351 \\
I am familiar with my group members' areas of expertise. & 5.333 \\
This was an interesting exercise. & 5.196 \\
My group members are familiar with my areas of expertise. & 5.175 \\
My group did not list important areas of my expertise. & 4.764 \\
I am confident that this system gives me new information. & 4.696 \\
This was a useful exercise. & 4.679 \\
I am confident that this system gives me good information. & 4.643 \\
I am willing to incorporate output from this system into my decision making. & 4.607 \\
I would be more comfortable with my group's tags if the tags were not anonymous. & 3.298 \\
\hline
Scale & Average Rating \\
\hline
Data Quality & 4.709  \\
Effort Expectancy & 4.670 \\
Result Demonstrability & 4.299 \\
Facilitating Conditions & 4.250 \\
Performance Expectancy & 3.836 \\
Relative Advantage & 3.742 \\
Anxiety (reverse coded) & 3.036 \\
\hline
\end{tabular}
}
\label{surveyresults}
\end{table}

All but one of the original items scored with mild to strong agreement. The 
highest ratings of agreement were received by the statements regarding comfort 
and familiarity of the group members with one another's areas of expertise. 
Additionally, nearly all participants rated this to be an interesting exercise.

Slightly lower ratings were received by the items regarding the results of 
the exercise. The participants believed the system gave them somewhat good and new
information that they found useful. They also thought the system 
did not necessarily gather all the important areas of their expertise and that 
they would not necessarily use the information to help them make decisions 
moving forward.

Being averages, the aggregate scales are relatively mild and all fit between 
3 and 5, straddling the Neutral rating. However, they showed similar results to 
the original items.

At the top of the list, the participants believed this exercise
provided good data quality and was easy to use and clear to understand. 
The participants rated the items regarding the results of the exercise and its 
fit within their organization slightly higher than neutral.

\section{Discussion}

\subsection{H1}

The data supports the view that this method provides 
a baseline for concluding that a group's opinion about a person's areas of 
expertise can give good information. A consensus appeared, was agreed to 
by the individual being labeled, and somewhat converged over time as the 
language and norms of the group were negotiated in a shared space. 

This finding comes with the caveat that the participants knew one another well 
enough or had enough experience with one another to feel the data being 
provided was of good enough quality. When conducted outside of well-known 
groups, this finding may not hold as both participant identity and the promise 
of future interactions are not as strong.

\subsection{H2}

This hypothesis was found to be partially supported. Participants did 
have confidence in the system to collect and then report the type of information 
they were expecting it to report. They thought the data would be quality data 
and they trusted it for what it was. 

However, they did not report that the trust in the data carried over to 
increased trust in the other participants. The study design forced the group 
members to already be acquainted with one another and have existing work- 
ing relationships. This means that the participants began the study with a 
fairly high degree of trust. This study provided no support for the idea that 
participants' trust levels increased because of the exercise. 

It would be interesting to ask a specific set of questions about colleague 
trust of a set of group members who were just beginning to work together or of 
group members who knew each other in a less formal environment than their 
salaried jobs.

\section{Conclusion}

Throughout this work, there was a sense that a kind of signaling was happening among members of the group, but implicitly through the data.  The participants did not report they brought the ongoing results of this study into their face-to-face conversations.

In a social space, judgment from others is ever-present and constant.  By bringing that sense of evaluation out in the open and working together, a new collective artifact was produced that changed over time.  This artifact held importance in the minds of the participants as they were looking squarely at the group's judgment of their skillsets and areas of expertise.  Their value to the group was laid bare in some sense, made explicit.

The participants reported the first round was the hardest - tough on the psyche - but also exciting, and potentially rewarding.  They also reported ongoing concern about being pigeon-holed for what they have already done or already shared.   Perhaps this type of public judgment encourages exploration and continued learning on the behalf of the individual.  Perhaps it discourages sharing in the longer term.

Overall, this research has provided insight into how familiar groups of individuals in the workplace can understand what their colleagues think of their areas of expertise.  This work has shown that, with simple keywords, group members can convey the salient areas of expertise of their colleagues to a degree that is deemed ``similar'' and of ``high quality'' by both third parties and those being evaluated.

Identity formation and negotiation is alive and well, and this research fits within the frames drawn by \cite{Goffman1959The-Presentation-of-} and \cite{Tajfel1986The-social-identity-} and furthered by \cite{boyd2002Faceted-Id/entity:-M,boyd2008Taken-Out-of-Context}.  We perform and we understand ourselves in part by understanding the reflections that come back to us from others \cite{Marchionini2009Information-Concepts}.

In a fast-moving networked workplace, this ability to gain insight into the knowledge of others with a simple trustable lookup may prove valuable.  Tapping into the collective understanding and distilled opinion of those around us could be a useful tool or sanity check against both direct and indirect individual claims of expertise.  Equally, it could serve as a weapon against misplaced modesty, allowing us to collectively reward those who deserve to be given credit when credit is due.

What remains an open question is whether this type of collective opinion mapping works in an environment beyond the walls of the relatively small, trusted workplace, where people know one another (stable identity) and have many incentives to behave and only say positive, professional things about one another (``the shadow of the future'' \cite{Axelrod1984The-Evolution-of-Coo}).

I hope to begin answering this larger question soon with work based on the open internet.

We now live in an ever-shrinking world of always-on connectivity and powerful communication devices.  Since these devices are two-way, they provide a voice (and a distribution platform) to millions who, prior, have never had a voice.  This is a remarkable achievement and serves as a testament to the incredible advance of technology and our collective striving for equality with regards to opinions and freedom of speech.  However, with monumental increases in the number of voices and opinions being shared, we demand a requisite increase in the power of tools to help us filter all this newfound information.  We need good knobs to help us determine where to direct our always-limited and increasingly precious amount of attention.

The freedom to listen to anyone has to be balanced with the practicality of not being able to listen to everyone.  We need tools that help us serve both of these needs, albeit not at the same time.  The tools need to be flexible enough to let us listen to whomever, whenever and wherever we want, and to reserve the right to change our minds at a later time.

Finding good sources of information is hard.  Knowing whom to listen to when the subject matter is beyond one's personal experience is a daunting and important problem, but one that can be reduced to an engineering problem with the right approach.

\bibliography{terrell}  
%

\end{document}